\documentclass[aps,prl,twocolumn,10pt,superscriptaddress,nofootinbib]{revtex4-2}

\usepackage{amsmath,amssymb,amsfonts,braket,bookmark}
\usepackage{xcolor}
\usepackage{hyperref}
\usepackage[normalem]{ulem} 
\hypersetup{colorlinks=true,linkcolor=blue,citecolor=blue,urlcolor=blue}
\usepackage{shuffle}
\newcommand{\cl}[1]{\mathcal{#1}}

\usepackage{tcolorbox}
\definecolor{airforceblue}{rgb}{0.36, 0.54, 0.66}
\definecolor{azure}{rgb}{0.0, 0.5, 1.0}
\newtcolorbox{tdbox}{colback=airforceblue!40!white,colframe=azure!90!black} 
\newcommand{\td}[1]{
	\if\notesOn1
	\begin{tdbox}
		#1
	\end{tdbox}
	\fi
}

\def\notesOn{1}
\renewcommand{\[}{\begin{equation}\begin{aligned}}
\renewcommand{\]}{\end{aligned}\end{equation}}
\begin{document}

\title{Learning the S-matrix from data: \\ Rediscovering gravity from gauge theory via symbolic regression}

\author{Nathan Moynihan}
\email{n.moynihan@qmul.ac.uk}
\affiliation{Queen Mary, University of London, Mile End Road, London E1 4NS, UK}

\date{\today}

\begin{abstract}
We demonstrate that modern machine-learning methods can autonomously reconstruct several flagship analytic structures in scattering amplitudes directly from numerical on-shell data. In particular, we show that the Kawai--Lewellen--Tye (KLT) relations can be rediscovered using symbolic regression applied to colour-ordered Yang--Mills amplitudes with Mandelstam invariants as input features. Using standard feature-selection techniques, specifically column-pivoted QR factorisation, we simultaneously recover the Kleiss--Kuijf and Bern--Carrasco--Johansson (BCJ) relations, identifying a minimal basis of partial amplitudes without any group-theoretic input. We obtain the tree-level KLT relations with high numerical accuracy up to five external legs, using only minimal theoretical priors, and we comment on the obstacles to generalising the method to higher multiplicity. Our results establish symbolic regression as a practical tool for exploring the analytic structure of the scattering-amplitude landscape, and suggests a general data-driven strategy for uncovering hidden relations in general theories. For comparison, we benchmark this general approach with a recently introduced neural-network based method.

\end{abstract}
\maketitle

\section{Introduction}
Machine learning has become an increasingly useful tool in the natural sciences, and recently there has been a growing interest in applying ML techniques to problems in formal theoretical physics \cite{doi:10.1142/q0404}. Outside of physics, the development of deep neural networks (DNNs) has been extremely rapid, and these models have been shown to be capable of performing a variety of tasks, from natural language processing to image recognition \cite{doi:10.1142/q0404}. That being said, the progress in applying these models to problems in theoretical physics has been somewhat slower, and this is in part due to the fact that deep neural networks have an \textit{interpretability problem}: they are excellent at making predictions, but the underlying mechanism by which they do so is often opaque. While it would undoubtedly be useful to have a DNN that could predict the numerical result of some experiment with better accuracy than our existing models, it's hard to see how this would constitute a ``discovery'' in the sense that we usually mean, since we would have no better understanding of the underlying physics than we did before \cite{rudin_stop_2019,Makke:2025vmd}.

Instead of simply outputting a prediction, \textit{symbolic} machine learning seeks to express learned relationships in terms of physically interpretable expressions. In particular, symbolic regression (SR) is a method for inferring algebraic, human-readable symbolic expressions from numerical data, making it a particularly appealing tool for theoretical physics. There have been several notable applications of SR in discovering physical laws of nature, for example AI-Feynman \cite{Udrescu:2019mnk} was able to rediscover $\mathcal{O}(100)$ well-known physical laws from synthetic data, such as the equations of motion for a double pendulum and the Hamiltonian for a two-body gravitational system. A recent application of SR was to use astronomical data from NASA to rediscover Newton's law of gravitation and all the planets' masses \cite{Lemos:2022cdj}. While these examples involve rediscovering empirical laws governing real, observable phenomena, it is interesting to consider whether SR can be used to probe more formal algebraic structures in quantum field theory.

Scattering amplitudes are not themselves directly observable, and therefore do not obey ``natural'' laws in the usual sense. That said, they are highly constrained by general physical principles, for example locality, unitarity, gauge invariance, little-group scaling, dimensional analysis etc. This makes them an especially attractive target for interpretable machine learning: exact analytic results are available as benchmarks, the underlying expressions exhibit striking hidden simplicity, and they play a dual role as both formal theoretical objects and building blocks for collider phenomenology and gravitational wave physics. The rich analytic structure of scattering amplitudes has been studied extensively over the past few decades, leading to a number of important discoveries, such as the BCFW recursion relations \cite{Britto:2005fq}, the CHY formalism \cite{Cachazo:2013hca}, and the double-copy relations between gauge and gravity theories \cite{Bern:2008qj}. Very recent work used a large language model to help reveal subtle structure in gauge theory, conjecturing a closed-form expression for single-minus tree-level gluon amplitudes and showing that these can be non-zero in restricted kinematic regions \cite{Guevara:2026smg}.

In this work, we take the Kawai--Lewellen--Tye (KLT) relations as a guiding target, and ask whether symbolic regression (specifically \textsc{pysr} \cite{cranmer_interpretable_2023}) can rediscover them directly from numerical amplitude data alone. The first step in any data-driven analysis is a completely standard feature-selection pass, and for us this will take the form of column-subset selection via column pivoted QR (CPQR) factorisation \cite{CHAN198767}. This is an interpretable form of feature selection to prune redundant colour orderings and kinematic invariants, and as we will see the familiar KK and BCJ relations then emerge automatically as the linear redundancies removed by this compression step. We then ask the important questions: what physical priors are necessary, and where do the limitations arise?

\section{Setup: amplitudes, relations and data}
\subsection{Scattering amplitudes of massless particles}
We will consider massless boson scattering amplitudes in four spacetime dimensions, where the external states are characterized by their four-momenta $p_i^\mu$ and helicities $h_i$. The momenta are null vectors, $p_i^2 = 0$, and satisfy momentum conservation, $\sum_{i=1}^n p_i^\mu = 0$. There are several ways to encode the kinematic data of massless particles (spinor-helicity variables, polarization vectors etc), but the precise construction will not concern us here. What will be important are the general properties of amplitudes in four-dimensions. For massless particles, little-group scaling implies that amplitudes transform as
\[
\cl{A}_n(\ldots, t_i\lambda_i,t_i^{-1}\tilde{\lambda}_i \ldots) = t_i^{-2h_i} \cl{A}_n(\ldots, \lambda_i,\tilde{\lambda}_i, \ldots),
\]
for each external leg $i$, under a little group transformation of the form $\Lambda^\mu_{~\nu}p^\nu = p^\mu$. This means that amplitudes have a definite helicity weight, along with a definite mass dimension, which in four-dimensions is given by $[\cl{A}_n] = 4-n$, where $n$ is the number of external legs. Finally, massless amplitudes typically depend on Mandelstam invariants $s_{ij} = 2 p_i \cdot p_j$. These invariants are not all independent, however, since they satisfy momentum conservation, $\sum_{j\neq i} s_{ij} = 0$, for each $i$. For $n$-point scattering, there are $\frac12 n(n-3)$ independent Mandelstam invariants. There are also dimensionally-dependent identities among the invariants arising from the vanishing of the Gram determinant for $n\geq 6$ in four dimensions. We will consider only the dimension-agnostic form in what follows.
\subsection{Scattering Amplitudes in Yang-Mills}
At tree-level, scattering amplitudes of pure gluons in Yang-Mills can be computed using Feynman diagrams, BCFW recursion relations \cite{Britto:2005fq}, or the CHY formalism \cite{Cachazo:2013hca}. What matters for us is that gluon amplitudes can always be decomposed into colour-ordered partial amplitudes $A_n$, which are gauge-invariant functions of the external momenta and helicities, 
\[
\cl{A}_n = \sum_{\sigma \in S_n/Z_n} \text{Tr}(T^{a_{\sigma(1)}} \cdots T^{a_{\sigma(n)}}) A_n(\sigma(1), \cdots, \sigma(n)),
\]
where the colour generators obey 
\[
~[T^{a},T^b] =if^{abc} T^c,~~~~~\text{Tr}(T^a T^b) = \delta^{ab},
\]
and the $A_n(\sigma(1), \cdots, \sigma(n))$ are the partial amplitudes which contain all of the kinematic information. Naively, for $n$ legs there are $n!$ partial amplitudes, however there are a number of important redundancies to consider. Firstly, the colour traces satisfy cyclic symmetry, $\text{Tr}(T^{a_1} T^{a_2} \cdots T^{a_n}) = \text{Tr}(T^{a_2} \cdots T^{a_n} T^{a_1})$, and therefore the partial amplitudes are cyclically symmetric, $A_n(1,2,\ldots,n) = A_n(2,\ldots,n,1)$, reducing the number of partials to $(n-1)!$. They also obey reflection symmetry, $A_n(1,2,\ldots,n) = (-1)^n A_n(n,n-1,\ldots,1)$, further reducing their number to $\frac12(n-1)!$. The partial amplitudes themselves also obey important kinematic relations, constraining the number of independent partials even more. Firstly, the Kleiss-Kuijf (KK) relations \cite{Kleiss:1988ne}
\[
A_n(1,\alpha,n,\beta) = (-1)^{|\beta|} \sum_{\sigma \in \alpha \shuffle \beta^T} A_n(1,\sigma,n),
\]
where $\alpha$ and $\beta$ are ordered subsets of $\{2,3,\ldots,n-1\}$, and $\beta^T$ is the reverse ordering of $\beta$, while $\sigma$ runs over all shuffles of $\alpha$ with $\beta^T$. This reduces the number of independent colour-ordered amplitudes from $\frac12(n-1)!$ to $(n-2)!$. Secondly, the fundamental Bern-Carrasco-Johansson (BCJ) \cite{Bern:2008qj} relations
\[
\sum_{i=2}^{n-1}\left(\sum_{j=i+1}^n s_{2j}\right) A_n(1,3, \ldots, i, 2, i+1, \ldots, n-1, n)=0 .
\]
takes the number of independent partials from $(n-2)!$ to $(n-3)!~$. Importantly, these relations are all linear in the partial amplitudes, and the relations above allow us to fix $3$ legs, say legs 1, $n-1$ and $n$, and express all other partial amplitudes in terms of the $(n-3)!$ independent amplitudes of the form $A_n(1,\alpha,n-1,n)$, where $\alpha$ runs over all permutations of $\{2,3,\ldots,n-2\}$.
\subsection{Gravity via the KLT Relations}
Tree-level scattering amplitudes of pure gravitons in General Relativity can also be computed using Feynman diagrams, BCFW recursion relations \cite{Britto:2005fq}, or the CHY formalism \cite{Cachazo:2013hca}. The important properties for us are that graviton amplitudes are fully permutation symmetric, and that they can be expressed in terms of gluon amplitudes via the Kawai-Lewellen-Tye (KLT) relations \cite{Kawai:1985xq}. The KLT relations state that tree-level graviton amplitudes can be expressed as a sum over products of colour-ordered gluon amplitudes, weighted by a kinematic kernel $S[\alpha|\beta]$, a pure function of Mandelstam invariants. For $n$-point tree-amplitudes, the KLT relations are
\begin{widetext}
\[
\cl{M}_n = (-1)^{n+1}\sum_{\alpha,\beta \in S_{n-3}}A_n(1,\alpha,n-1,n) S[\alpha|\beta] A_n(1,\beta,n,n-1),
\]
\end{widetext}
where $\alpha$ and $\beta$ run over all permutations of the set $\{2,3,\ldots,n-2\}$, and the KLT kernel $S[\alpha|\beta]$ is given by
\[
S[\alpha|\beta]=\prod_{k=2}^{n-2}\left(s_{1,\alpha_k}+\sum_{\ell=2}^{k-1}\theta(\alpha_k,\alpha_\ell|\beta)\,s_{\alpha_k\alpha_\ell}\right),
\]
where $\theta(i_\alpha,i_j|\beta) = 1$ iff $i_\alpha$ appears before $i_j$ in the ordering $\beta$, and zero otherwise.
This is best understood by way of some examples, which will also serve as useful benchmarks later on. At 4-point, $\alpha = \beta = 2$, and therefore $S[2|2] = s_{12}$, and the KLT relation is simply
\[
\cl{M}_4 = - s_{12} A_4(1,2,3,4) A_4(1,2,4,3).
\] 
At 5-point, $\alpha,\beta \in \{2,3\}$, and we have to consider all permutations. The KLT kernel is given by
\[
S[2,3|2,3] &= s_{12}s_{13}\\
S[2,3|3,2] &= s_{12}(s_{13}+s_{23})\\
S[3,2|2,3] &= s_{13}(s_{12}+s_{23})\\
S[3,2|3,2] &= s_{13}s_{12},
\]
and so the corresponding gravity amplitude is 
\[
\cl{M}_5 =&~~ s_{12}s_{13} A_5(1,2,3,4,5) A_5(1,2,3,5,4) \\
&+ s_{12}(s_{13}+s_{23}) A_5(1,2,3,4,5) A_5(1,3,2,5,4) \\
&+ s_{13}(s_{12}+s_{23}) A_5(1,3,2,4,5) A_5(1,2,3,5,4) \\
&+ s_{13}s_{12} A_5(1,3,2,4,5) A_5(1,3,2,5,4).
\]
With these examples in mind, we now turn to the task at hand: can we discover the KLT relations from numerical data alone?

\section{Linear structure from data}
Our overall strategy is to treat the KLT relation as the destination, assuming only the $\frac12(n-1)!$ colour-ordered Yang--Mills amplitudes, along with Mandelstam invariants, as features. We then compress the raw feature space using off-the-shelf linear feature selection, and then use symbolic regression on our reduced feature set to reconstruct the gravity amplitude (which we compute via Hodges formula \cite{Hodges:2011wm,Hodges:2012ym}), with the KK and BCJ relations appearing along the way as the linear relations exposed by the feature reduction step.

\subsection{Data and Kinematics}
We work at tree-level, in four dimensions, with massless external states satisfying all-outgoing kinematics. Random on-shell kinematics are generated in the centre-of-mass frame by sampling $2\rightarrow(n-2)$ momenta, satisfying momentum conservation. We work with exact rational kinematics initially, although for the symbolic regression step we convert to 64-bit floating-point numbers, and we impose the on-shell and transversality conditions to within a numerical tolerance of $\cl{O}(10^{-16})$.

Spinor–helicity variables are constructed from the sampled four-vectors, and colour-ordered MHV gluon amplitudes are computed using the Parke-Taylor formula \cite{Parke:1986gb} for the appropriate orderings, with the MHV gravitational targets computed similarly. Since generic spinor-helicity variables in $(3,1)$ signature are complex, the amplitudes are complex-valued. To avoid doubling the feature space by treating real and imaginary parts separately, we analytically continue to $(2,2)$ signature, where spinor products and amplitudes are real.

For an $n$-point tree amplitude of massless gluons in a fixed helicity configuration, we take ${\cl{O}_1,\ldots,\cl{O}_N}$ to denote a set of $N$ colour orderings. Evaluating each ordering at $M$ random kinematic points yields an $M\times N$ amplitude data matrix
\[
    \mathbf{A} = \begin{pmatrix}
      A_1^{(1)} & A_1^{(2)} & \cdots & A_1^{(N)} \\
      A_2^{(1)} & A_2^{(2)} & \cdots & A_2^{(N)} \\
      \vdots    & \vdots    &        & \vdots    \\
      A_M^{(1)} & A_M^{(2)} & \cdots & A_M^{(N)}
    \end{pmatrix}
\]
where $A_k^{(i)} \equiv A_n(\cl{O}_i)$ evaluated at the $k$-th phase space point. In parallel, we construct an $M\times P$ matrix of Mandelstam invariants,
\[
\mathbf{S} = \begin{pmatrix}
      s_{12}^{(1)} & s_{13}^{(1)} & \cdots & s_{n-1,n}^{(1)} \\
      s_{12}^{(2)} & s_{13}^{(2)} & \cdots & s_{n-1,n}^{(2)} \\
      \vdots    & \vdots    &        & \vdots    \\
      s_{12}^{(M)} & s_{13}^{(M)} & \cdots & s_{n-1,n}^{(M)}
  \end{pmatrix},
\]
and an $M\times 1$ vector of gravitational targets,
\[
\mathbf{M} = \begin{pmatrix}
      M_1 \\
      M_2 \\
      \vdots    \\
      M_M
  \end{pmatrix}.
\]
For numerical stability we discard phase-space points close to singular kinematics; concretely, we reject any point for which some Mandelstam invariant satisfies $|s_{ij}| < \delta$, for some small cutoff $\delta$ of order $\cl{O}(10^{-10})$, say. We also rescale all invariants and amplitudes by some values in the generated sample,
\[
\tilde{s}_{ij} = \frac{s_{ij}}{S},~~~~~\tilde{A}_n = \frac{A_n}{A},~~~~~ \tilde{M}_n = \frac{M_n}{M},
\]
where we either take $S = A = M = 10^k$ to be some fixed scale with integer $k\geq 0$, or we take $S = \text{median}(\{s_{ij}\})$, $A = \text{median}(\{|A_n|\})$ and $M = \text{median}(\{|M_n|\})$. This ensures that all features and targets are dimensionless $\cl{O}(1)$ numbers, which again improves numerical stability. The final dataset will eventually be a matrix of random phase-space points, with columns corresponding to the colour-ordered gluon amplitudes, the Mandelstam invariants, and the target gravitational amplitude
\[
\mathbf{X} = \left(\mathbf{S}'~|~\mathbf{A}'\right),~~~\mathbf{y} = \mathbf{M}
\]
where $\mathbf{S}'$ and $\mathbf{A}'$ are the potentially pruned sets of invariants and amplitudes after feature selection.
\subsection{Structure discovery via CPQR}
From a physics perspective, the space of colour-ordered amplitudes and Mandelstam invariants is highly redundant. At tree level, this redundancy is encoded in familiar relations such as Kleiss--Kuijf (KK) \cite{Kleiss:1988ne} and Bern--Carrasco--Johansson (BCJ) \cite{Bern:2008qj} relations among amplitudes, as well as momentum-conservation identities for invariants. In this section, however, we deliberately avoid imposing any of these structures \emph{a priori}. Instead, we will treat amplitudes and invariants as raw numerical features and ask what linear relations can be inferred using standard, interpretable linear-algebraic diagnostics. The motivation here is purely pragmatic: for symbolic regression of the KLT relations, we need a compact, non-redundant set of inputs, and CPQR provides an interpretable way to obtain it. Identifying KK/BCJ is simply a check on what was removed.

The guiding observation is simple: if a set of sampled features obeys exact linear relations, then the corresponding data matrix is rank-deficient, up to numerical tolerance, and we should perform a dimensional reduction. To extract these relations, we will use \emph{column-pivoted QR (CPQR) factorisation} to select a minimal, physically interpretable subset of \emph{original} columns that spans the data.

\subsubsection{Column subset selection and the Kleiss--Kuijf relations}
The gold standard in dimensional reduction in machine learning is principal component analysis (PCA), which efficiently reveals low-dimensional structure in a numerical dataset. However, the basis vectors in the set are typically linear combinations of the original features, which doesn't lend itself to interpretability, for example a PCA mode could mix objects with different mass dimension, say an amplitude added to a Mandelstam variable, which has no sensible physical meaning. Since our goal is interpretable structure discovery, we will instead use CPQR factorisation, which selects a subset of \emph{existing} columns that approximates the full column space to within a chosen tolerance.

We apply this procedure to the amplitude and invariant matrices $\mathbf{A}$ and $\mathbf{S}$ defined in the previous subsection. Here, rows correspond to random phase-space points and columns correspond to distinct features. In this language, the Kleiss--Kuijf statement that only $(n-2)!$ colour-ordered amplitudes are independent becomes the statement that the columns of $\mathbf{A}$ span an approximately $(n-2)!$-dimensional subspace of $\mathbb{R}^M$.

Concretely we compute
\[
\mathbf{AP} = \mathbf{QR},
\]
where $\bf{P}$ is a permutation matrix which performs the pivots, $\mathbf{Q}\in\mathbb{R}^{M\times N}$ is a semi-unitary matrix (with orthonormal columns), and $\bf{R}$ is an upper-triangular matrix of the form
\[
\bf{R} = \begin{pmatrix}
      R_{11} & R_{12}  \\
      0 & \epsilon R_{22}
  \end{pmatrix} = \begin{pmatrix}
      R_{11} & R_{12}  \\
      0 & 0
  \end{pmatrix} + \cl{O}(\epsilon),
\] 
A sharp drop in the diagonal entries of $\mathbf{R}$ identifies an effective numerical rank $k$, in which case the first $k$ columns of $\mathbf{AP}$ provide a basis for the column space of $\mathbf{A}$ to within our chosen tolerance $\epsilon$. This in turn implies that there are several linear relations among the columns of $\mathbf{A}$.

To simply perform symbolic regression, it is not a requirement that we know how the columns are related -- we only need a reduced set of independent features to feed into our regression algorithm. However, it is interesting to see what relations are actually being uncovered by CPQR, and so we will extract from the $R$ matrix by explicitly constructing the nullspace basis $\bf{N}$ as
\[
\bf{N} = \bf{P}\begin{pmatrix}
      -R_{11}^{-1}R_{12}  \\
      I
  \end{pmatrix},~~~~~\text{so that}~~~~~\bf{AN} \simeq 0.
\]
The columns of $\mathbf{N}$ therefore give rise to null relations among the sampled features. In algebraic language, such relations are \textit{syzygies} \cite{Kosower:2025inx}\footnote{We thank David Kosower for bringing our attention to this work, and for useful conversations on the topic.}: nontrivial coefficient vectors $\mathbf{c}(s_{ij})$ (viewed as polynomials/rational functions of the kinematic invariants) for which a linear combination of candidate features vanishes identically on the kinematic support. In the present subsection the extracted syzygies are degree-0 in the invariants, since they have (approximately) constant coefficients, and they reproduce the Kleiss--Kuijf relations \cite{Kleiss:1988ne}.

For example, using $\cl{O}(n^3)$ random phase space points we find that the effective rank of $\mathbf{A}$ is $k=(n-2)!$, with a pronounced multi-order-of-magnitude gap in the diagonal of $\mathbf{R}$ after the first $k$ pivots. 

The resulting null relations (syzygies) $\mathbf{A}\mathbf{N}\simeq 0$ contain the celebrated KK relations \cite{Kleiss:1988ne} up to numerical error. We emphasise that this extraction requires no group-theoretic input: the relations emerge from numerical data and linear algebra alone. Applying the same CPQR analysis to $\mathbf{S}$ similarly recovers the expected momentum-conservation identities among Mandelstam invariants.
\subsubsection{Joint feature-space structure and the Bern--Carrasco--Johansson relations}
After pruning $\mathbf{A}$ and $\mathbf{S}$ using CPQR, the simplest natural input set for symbolic regression is the concatenated matrix
\[
\mathbf{X} = \left(\mathbf{S}'~|~\mathbf{A}'\right),
\]
where $\mathbf{S}'$ and $\mathbf{A}'$ denote the surviving invariant and amplitude columns. Although this feature set is sufficient in principle to express the gravitational targets, it requires the symbolic regression algorithm to discover \emph{a priori} that amplitudes and Mandelstam invariants must first be multiplied to form objects of the appropriate mass dimension and little-group weight. This is inefficient, as it greatly enlarges the search space.

A standard remedy in symbolic regression is to include composite features suggested by our knowledge of the problem. In our case, we use only minimal physical information about the target sector (mass dimension and little-group weights) to build a composite feature library, for example
\[
\{\,s_{ij}\,A_\alpha\,\},\qquad \{\,s_{ij}s_{kl}\,A_\alpha\,\},\qquad \{\,s_{ij}\,A_\alpha A_\beta\,\},\ \ldots
\]
where $A_\alpha$ denotes a surviving colour-ordered amplitude column.

After forming these products, we apply CPQR again to prune the enlarged feature set. Since the number of candidate composites grows rapidly, the precise truncation of the library is ultimately a practical choice. Importantly, the subsequent pruning and regression remain fully data-driven; our only theoretical input is the choice to include composites compatible with the target mass dimension and little-group weights.

Applying CPQR to the composite feature matrix also provides a natural interpretation in terms of \emph{degree-1 syzygies} \cite{Kosower:2025inx}. Concretely, consider the set of features with columns of the form $(s_{ij}A_\alpha)$, arranged into a matrix $\mathbf{A}^{(1)}$. A null vector $\mathbf{N}$ satisfying
\[
\mathbf{A}^{(1)}\mathbf{N}\simeq 0
\qquad\Longleftrightarrow\qquad
\sum_{\alpha,\,ij} c_{\alpha,ij}\,s_{ij}\,A_\alpha \simeq 0
\]
can be regrouped as
\[
\sum_{\alpha}\Big(\sum_{ij} c_{\alpha,ij}\,s_{ij}\Big)A_\alpha \simeq 0.
\]
Thus, a constant-coefficient null relation in the composite space corresponds to a relation among amplitudes whose coefficients are \emph{linear polynomials} in the Mandelstams, i.e.\ a degree-1 syzygy in the invariants, which of course contain the Bern--Carrasco--Johansson relations \cite{Bern:2008qj} directly from numerical data and linear algebra, without imposing amplitude identities by hand. In other words, CPQR automatically discovers that certain linear combinations of amplitudes, weighted by Mandelstam polynomials, vanish identically---precisely the BCJ relations.

However, it should be noted that the nullspace basis returned by CPQR is not unique, and the resulting relations are often not presented in the canonical BCJ form, but rather as generic linear combinations of a BCJ basis (and often overcomplete). At low points this can be massaged into the familiar representation, but at higher points it becomes impractical, and it would be interesting to explore algorithmic ways to extract the canonical BCJ basis directly. We can, in principle, extend this to higher-degree syzygies by including higher-order composites, e.g. $\{s_{ij}s_{kl}A_\alpha\}$ (see \cite{Kosower:2025inx} where this has been done). However, for purely symbolic regression purposes this tends to reduce the flexibility of the feature set: the library blows up and is full of near-duplicates, and pruning then leaves you with ``pre-multiplied'' high-degree features that are less convenient for building simple formulas.
\section{Symbolic reconstruction}
We now turn to learning about amplitudes using symbolic regression. Given a target $y$, the two central design choices are the input feature set ${x_i}$ and the operator/function class over which the search is performed. For tree-level amplitudes, the relevant building blocks are Lorentz scalars (which could be spinor-helicity objects) and rational functions thereof. We therefore consider polynomial/rational expression spaces generated by
\[
\{+,-,\times,/\},
\]
and in practice often implement division via a unary \textbf{inv} operator, $x \mapsto 1/x$, for improved search stability.

A crucial practical point is that the symbolic search space grows rapidly with both the number of input features and the allowed operator set. The CPQR pipeline therefore serves a dual purpose: it is not only a discovery tool for linear amplitude identities, but also a mechanism for reducing the combinatorial burden on symbolic regression.

\begin{figure}[t]
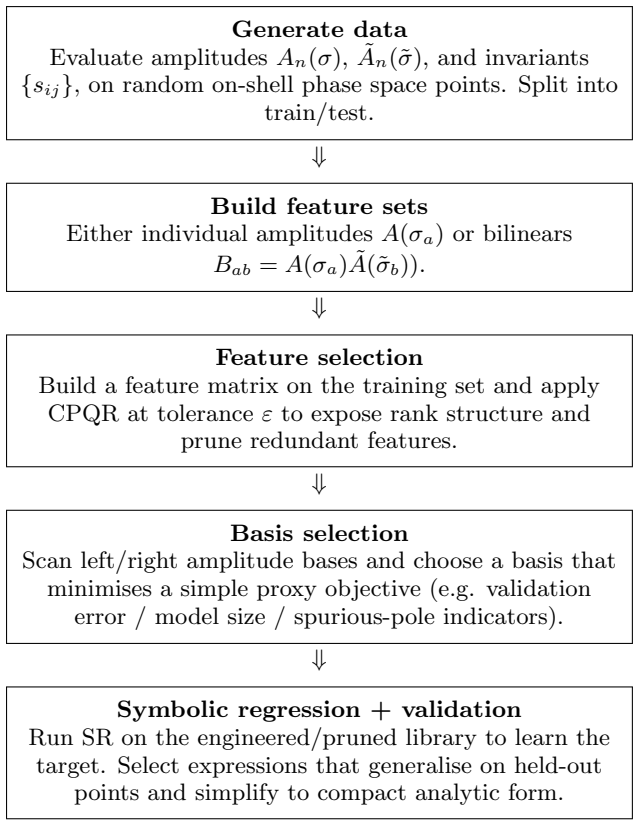

\centering
\setlength{\fboxsep}{5pt}
\renewcommand{\arraystretch}{1.15}
\begin{tabular}{c}
\fbox{\parbox{0.92\linewidth}{
\textbf{Generate data}\\
Evaluate amplitudes \(A_n(\sigma)\), \(\tilde A_n(\tilde\sigma)\), and invariants \(\{s_{ij}\}\), on random on-shell phase space points. Split into train/test.
}}\\[4pt]
$\Downarrow$\\[4pt]
\fbox{\parbox{0.92\linewidth}{
\textbf{Build feature sets}\\
Either individual amplitudes $A(\sigma_a)$ or bilinears \(B_{ab}=A(\sigma_a)\tilde A(\tilde\sigma_b)\)).
}}\\[4pt]
$\Downarrow$\\[4pt]
\fbox{\parbox{0.92\linewidth}{
\textbf{Feature selection}\\
Build a feature matrix on the training set and apply CPQR at tolerance \(\varepsilon\) to expose rank structure and prune redundant features.
}}\\[4pt]
$\Downarrow$\\[4pt]
\fbox{\parbox{0.92\linewidth}{
\textbf{Basis selection}\\
Scan left/right amplitude bases and choose a basis that minimises a simple proxy objective (e.g. validation error / model size / spurious-pole indicators).
}}\\[4pt]
$\Downarrow$\\[4pt]
\fbox{\parbox{0.92\linewidth}{
\textbf{Symbolic regression + validation}\\
Run SR on the engineered/pruned library to learn the target. Select expressions that generalise on held-out points and simplify to compact analytic form.
}}\\
\end{tabular}
\caption{Schematic of the pipeline used to rediscover KLT-type relations from numerical samples.}
\label{fig:pipeline}
\end{figure}

Finally, we note that further performance gains can be achieved by modest, physics-motivated feature engineering. We can use our prior amplitude knowledge, for example that they have definite mass dimension and little-group weights, and use this to build composite physics-aware inputs with the correct overall scaling for the target sector. For instance, at four points in the MHV sector, $[\cl{M}_4]=0$ and the little-group weights are
\[
(-2h_1,-2h_2,-2h_3,-2h_4) = (+2,+2,-2,-2),
\]
so it is natural to prioritise feature combinations that respect these constraints. In practice, the most effective strategy balances two competing considerations: the \emph{expressivity} of the engineered inputs against the \emph{complexity} of the resulting symbolic search space. In practice, the feature set we pick is bounded: beyond bilinear in the gluon amplitudes, any resulting expression will have the wrong little-group scaling, and beyond $s_{ij}^{n-3}$ in Mandelstam invariants the mass dimension will be incorrect. The full pipeline is summarised in Fig.~\ref{fig:pipeline}, which we will now apply to the Parke-Taylor and KLT problems in turn.

\subsection{Parke-Taylor Amplitudes}
As a warm-up, and as a benchmark en route to the KLT problem, we consider a landmark result in the study of scattering amplitudes: the Parke-Taylor formula \cite{Parke:1986gb} for tree-level MHV gluon amplitudes. The Parke-Taylor formula gives a remarkably simple closed-form expression for the $n$-point MHV amplitude with two negative-helicity gluons (say legs $1$ and $2$) and $(n-2)$ positive-helicity gluons
\[
A_n[1^-,2^-,3^+,\ldots, n^+] = \frac{\braket{12}^4}{\braket{12}\braket{23}\cdots \braket{n1}}
\]
This should be contrasted with the complexity of the Feynman-diagram expansion, which grows factorially with $n$. We ask: can symbolic regression rediscover the Parke-Taylor formula directly from numerical data? While this may seem like a simple exercise, it has enough non-trivial aspects to serve as a good test of the method. The Parke-Taylor formula is a rational function of spinor products, and cannot be expressed as a polynomial in Mandelstam invariants. At the same time, it is a relatively simple expression, and therefore should be within reach of symbolic regression. It therefore serves as a useful benchmark to set our expectations: it has several known physical properties that can be used to guide our exploration, and so we will start with minimal priors and gradually add structure to gauge their effect.

Our target is the $n$-point MHV amplitude $A_n[1^-,2^-,3^+,\ldots, n^+]$, evaluated on $M$ random phase space points, forming an $M\times 1$ target vector $\mathbf{A}^\text{target}$, perhaps generated by Feynman diagrams, with the full set of spinor brackets as features:
\[
\{\braket{12},[12],\braket{13},[13],\cdots,~|~\mathbf{A}^\text{target}\}.
\]
For $n$ particles, there are $n\choose 2$ independent angle brackets and $n\choose 2$ independent square brackets, and so the feature space grows quadratically with $n$. We then run symbolic regression (SR) on this dataset, searching for an expression that fits the target amplitude. For $n=4$, SR is able to rediscover the Parke-Taylor formula with minimal difficulty, in $\cl{O}(10^2)$ seconds, although often in a peculiar form\footnote{For example \[A_4 = -\frac{[34]^2\braket{12}}{\braket{23}[23][12]}\]}. However, as we increase $n$, the search space grows \textit{rapidly}, and so SR struggles to find the correct expression within a reasonable time-frame --- $n=5$ requires $\cl{O}(10^{4})$ seconds, for example. The best way to improve this situation is to include additional priors, either to shrink the feature set, including only a subset of spinor-helicity variables for example, or the search space itself, by restricting to a particular set of operators/functions. We must be careful here, however, to not make assumptions based on the known Parke-Taylor form itself, since this would defeat the purpose of rediscovery.

At tree-level, we know that amplitudes are polynomials in spinor products along with poles corresponding to factorisation channels. We can't therefore really do better than $\{+,\times, \textbf{inv}\}$ without prior knowledge that Parke-Taylor happens to be a \emph{single} fraction, which would be cheating. We can help the algorithm along by performing some physics-inspired feature engineering, i.e. using known physical properties of the target amplitudes to bias the search space. Firstly, we turn to dimensional analysis, knowing that tree-level amplitudes have dimension $[\cl{A}_n] = 4-n$, we can enforce that this is as a constraint (this is built-in to \textsc{pySR} \cite{cranmer_interpretable_2023}), taking $[\braket{ij}] = 1$ and a dimensionless coupling. Secondly, it is well known that (tree-level) MHV gluon amplitudes in four dimensions admit a chiral representation: they can be written purely in terms of the undotted spinors $\lambda_i$, therefore depending only on angle brackets $\langle ij\rangle$, with the parity-conjugate $\overline{\text{MHV}}$ sector depending only on $[ij]$. This holomorphy is natural when Yang--Mills is treated perturbatively as an expansion around its self-dual sector, and it is manifest in twistor-space and light-cone/MHV-Lagrangian formulations where the relevant MHV vertices are explicitly holomorphic \cite{Chalmers:1996rq,Witten:2003nn,Mason:2005zm,Cachazo:2004kj,Mansfield:2005yd,Ettle:2006bw,Nair:1988bq}. Motivated by this known chiral structure of the tree-level MHV sector, we can optionally restrict ourselves to angle brackets only. This reduces the search space without specifying the cyclic denominator structure, resulting in the dataset
\[
\{\braket{12},\braket{13},\ldots ~|~ \mathbf{A}^\text{target}\},
\]
With these considerations alone, we find that SR rediscovers the Parke-Taylor formula at $n=4,5,6$ in a reasonable time, i.e. under $\mathcal{O}(10^3)$ seconds, to $\cl{O}(10^{-15})$ maximum relative error on a held out test set. This is an orders-of-magnitude speed up over the full feature set and unrestricted search space. Stronger priors, such as restricting to a single fraction or only using adjacent spinor brackets, speed the search up further, but are not physically well-motivated and are unnecessary up to $n=6$ anyway.
\subsection{Learning The KLT Relations}
Having benchmarked the method on the Parke-Taylor formula, we now turn to our main target: the KLT relations between tree-level gluon and graviton amplitudes, a clean target for bootstrap attempts \cite{Chi:2021mio}. Our core assumption is that the graviton amplitude can be expressed as a function of gluon amplitudes and Mandelstam invariants, i.e. $\cl{M}_n = f(A_n, s_{ij})$, with no assumption about the particular structure beyond what we can glean from physics. We continue to work in the MHV sector; at four and five points this captures all helicity configurations, while NMHV first appears at $n\geq 6$.

Our first task will be to construct a feature set suitable for symbolic regression. The most straightforward choice would be to simply take all colour-ordered gluon amplitudes $A_n(\cl{O}_i)$ and all Mandelstam invariants $s_{ij}$ as features. However, this leads to a combinatorial explosion in the size of the feature space, since the number of colour-ordered amplitudes grows as $(n-1)!$. That being said, we have already seen that our naive features are highly redundant, and so we can use the CPQR pipeline to prune both sets to a basis of independent features, and so we will end up with $(n-3)!$ independent amplitudes and $n(n-3)/2$ Mandelstams after CPQR, which should be manageable at least for small $n$. Again using the polynomial/rational function space generated by $\{+,\times,/\}$, it is obvious from the last section that the size of this feature space is a problem for symbolic regression, so we again use physics-motivated feature engineering to build composite features with the correct overall mass dimension and little-group scaling for the gravitational target. Firstly, since Mandelstams carry little-group weight zero, it must be the case that graviton amplitudes are bilinear in gluon amplitudes to get the correct little-group scaling. In $D=4$, scattering amplitudes have mass dimension $[\cl{A}_n] = 4-n = [\kappa^{n-2}\cl{M}_n]$, and $[\kappa] = -1$. The stripped graviton amplitude has mass dimension $[\cl{M}_n] = 2$, and so the kinematic function multiplying $A_n\tilde{A}_n$ must have dimension $2(n-3)$, i.e. be a homogeneous polynomial of degree $n-3$ in Mandelstams. In other words, the amplitude must be of the form 
\[
\cl{M}_n = \sum_{a,b}f_{ab}(s^{n-3}_{ij})A_n(\sigma_a)\tilde{A}_n(\tilde{\sigma}_b).
\]
The choice of orderings $\sigma_a$ and $\tilde{\sigma}_b$ has a significant effect on the complexity of the resulting expression, and especially the function $f_{ab}(s^{n-3}_{ij})$. Colour-ordered gluon amplitudes are not manifestly Bose-symmetric, because the colour-ordering selects \textit{planar} singularity channels only, having poles only in adjacent momenta. For example, the four-point amplitude $A_4(1,2,3,4)$ has simple poles in $1/s_{12}$ and $1/s_{23}$, but not in $1/s_{13}$, so symmetry under $1\leftrightarrow 3$ exchange is not manifest. Graviton amplitudes, on the other hand, must be fully Bose-symmetric, since gravitons are colourless and indistinguishable, and so we should expect graviton amplitudes to have poles in all channels, not just adjacent ones. Importantly, if the pair of chosen orderings $\sigma_a$ and $\tilde{\sigma}_b$ do not together cover all possible channels, then the functions $f_{ab}(s^{n-3}_{ij})$ \textit{must} introduce additional poles in Mandelstam invariants to ensure that the full gravitational amplitude is permutation-invariant. It should be noted, however, that even if the pair of orderings do cover all channels, there is no guarantee that $f_{ab}(s^{n-3}_{ij})$ will be free of spurious poles, since they can still appear provided they cancel in the full expression.

This is easy to see with an example at four-point. Choose the orderings $\sigma = (1,2,3,4)$ and $\tilde{\sigma} = (1,2,4,3)$, and consider the ansatz
\[
\cl{M}_4 = f(s^{n-3}_{ij})A_4(1,2,3,4)\tilde{A}_4(1,2,4,3),
\]
where $f(s_{ij})$ is taken, for the moment, to be a pole-free polynomial function. This choice is at least compatible with locality and factorisation: $A_4(1,2,3,4)$ has poles in $1/s_{12}$ and $1/s_{23}$, while $A_4(1,2,4,3)$ has poles in $1/s_{12}$ and $1/s_{24}$. Their product therefore has simple poles in $1/s_{12}$, $1/s_{23}$ and $1/s_{24}$, covering all possible channels at four-point, and eliminating the need for $f(s_{ij})$ to introduce any additional poles. However, the absence of a need for poles does not mean poles cannot be shifted into $f(s_{ij})$ by changing the partial-amplitude basis. Indeed, using the four-point BCJ relation, we can write
\[
\cl{M}_4 &= \sum_{a,b}\frac{f_{ab}(s^{n-3}_{ij})s_{23}}{s_{13}}A_4(1,2,3,4)\tilde{A}_4(1,2,3,4)\\
&= \sum_{a,b}f'_{ab}(s^{n-3}_{ij})A_4(1,2,3,4)\tilde{A}_4(1,2,3,4).
\]
All of this raises the question: which orderings should we choose when building our feature set? There is no obvious physical reason to choose one particular ordering over another, since they are all related by amplitude relations. We know from the CPQR/BCJ that there are $(n-3)!$ independent amplitudes at $n$-point, so many equivalent orderings exist. Since our goal is to rediscover compact analytic expressions, we would like the chosen orderings to bias the search toward minimal complexity, ideally avoiding spurious poles in $f_{ab}(s^{n-3}_{ij})$ if possible. In particular, using the same ordering in both copies tends to force $f(s_{ij})$ to introduce poles to reproduce missing channels, whereas using distinct orderings can distribute the physical poles across the two gauge-theory factors and permit simpler coefficient functions. Unfortunately, there is no physical principle that tells us which orderings are best in this regard, and so we must again let the data guide us. 

Our strategy in this case is to use a simple, computationally cheap decision-tree model to guide our choice of orderings. We construct candidate feature sets from different left-right ordering choices and use a fast decision-tree-based model to test how well each feature set approximates the gravitational target, ranking candidates by mean-squared error (MSE). If a particular ordering choice allows the target to be fit accurately using only a few features, this is strong evidence that the underlying analytic relation is comparatively simple in that basis, and so we can then use that ordering choice to build our feature set for symbolic regression, i.e. we build bilinears of the form

\[
B_{ab} = A_n(\sigma_a)\,\tilde A_n(\tilde\sigma_b),
\]
together with an appropriate set of Mandelstam invariants $\{s_{ij}\}$, chosen in the same way.
\subsubsection{Results and Scaling}\label{sec:klt_results}
We now have a pipeline for rediscovering the KLT relations, and we can apply it at four, five and six points to see how it performs. Success is counted as the rediscovery of a KLT-type relation within a fixed compute budget (we take 8 hours in the runs below), with success defined as matching the gravitational target to within numerical tolerance on a held-out test set (within $\cl{O}(10^{-16})$, and ideally producing a compact expression in terms of gluon amplitudes and Mandelstam invariants. The gravitational target is generated independently using Hodges’ formula \cite{Hodges:2011wm,Hodges:2012ym}, so as to avoid biasing the search toward the KLT form.

At four points the problem is essentially one-dimensional from the BCJ point of view: there is only \((n-3)!=1\) independent colour-ordered amplitude per copy, so the symbolic regression task is relatively straightforward. The pipeline rediscovers the expected relation with minimal difficulty, typically finding several equivalent expressions that agree within numerical tolerance, e.g.
\[
\mathcal M_4 = -s_{12}\,A_4(1,2,3,4)\,\tilde A_4(1,2,4,3).
\]
(As discussed above, different ordering choices can shift kinematic factors between the amplitudes and the coefficient function, so the same \(\mathcal M_4\) can look polynomial or rational depending on basis.)

At five points the situation is more interesting, since there are now multiple BCJ-independent orderings in each copy. Using the tree-based ordering-selection procedure outlined above, discovery within the allotted timeframe is frequently achieved, but the time-to-solution is highly sensitive to the particular left-right basis chosen. In some cases the correct relation is recovered in \(\mathcal O(10^3)\) seconds, while in others it takes \(\mathcal O(10^4)\) seconds, and for many basis choices the algorithm fails to converge within the 8-hour limit. Empirically, pairs of bases whose pole structure covers the physical factorization channels tend to yield lower MSE in the decision-tree analysis step, and then tend to enjoy substantially faster symbolic discovery. Conversely, “unfortunate” bases often lead to more complicated rational kernels with spurious poles and delicate cancellations, which dramatically slows the search, often failing in the allotted time. For example, one particularly simple such attempt (reproduced using seed 1096) yields the expression
\[
\cl{M}_5 = ~&(s_{35} + s_{23} - s_{14})s_{13}A_5(1,2,5,4,3)A_5(1,4,2,5,3)\\
&-s_{15}s_{23}A_5(1,2,3,4,5)A_5(1,4,2,3,5)
\]
in under a minute, which evaluates on a held-out test set with max error $\cl{O}(10^{-16})$.

At six points the problem becomes distinctly more challenging. The BCJ basis size grows to $(n-3)!=6$, so even before introducing any Mandelstam dependence there are $36$ bilinears $A(\sigma_a)\tilde A(\tilde\sigma_b)$ available. Dimensional analysis suggests that the corresponding kernel should involve a homogeneous polynomial of degree $n-3=3$ in Mandelstams (mass dimension $6$). Even if we restrict to a minimal independent set of nine invariants, the number of plausible composite features grows rapidly, and the resulting search space becomes prohibitively large. This is a simple result of combinatorics: the number of possible expressions already has Catalan-number growth in the expression size (counting distinct binary-trees), and once we also account for the allowed operators and number of features, the total number of candidate expressions blows up extremely fast. For an expression of complexity $m$ with $K$ features and 4 operators $\{+,-,*,/\}$, a back of the envelope estimate of the size of the search space is \cite{langdon:GPEM:gpconv,Tevfik4Op}
\[
\mathrm{Cat}_m 4^m K^{m+1} \sim \frac{16^m}{m^{3/2}\sqrt{\pi}}K^{m+1}.
\] 
This means that even a modest increase in the number of features $K$ can lead to an explosion in the search space, which is then further exacerbated by the fact that many of the features are strongly correlated, and that the 6pt KLT relation could have high complexity, on the order of $m\sim 500$ depending on the feature set used and how expanded the expression is.

There are several physically motivated biases we could introduce to try and tame this explosion. For example, we could use the 6pt syzygy relations discovered in \cite{Kosower:2025inx,Bjerrum-Bohr:2011jrh} to reduce the number of independent amplitudes (introducing spurious poles into $f(s_{ij})$, as discussed above), and we could also further restrict the space of Mandelstams by using the non-linear dimensionally-dependent identities in $D=4$. However, in practice this would still leave the search space far too large, and still we would find that regression fails to converge within the allotted time, even with aggressive feature engineering and basis selection. Concretely, the tree-based scan can still identify ``better'' bases according to our proxy objectives, but the subsequent symbolic regression step does not reliably rediscover the six-point KLT relation within the compute budget.

In short, the bottleneck at six points and beyond is a genuine combinatorial explosion in the effective feature space, made worse by the strong correlations among invariants and by basis choices that push rational structure (and spurious cancellations) into the learned function $f$. Put slightly differently: at five points we can often ``get away with'' a basis in which the kernel looks like a short polynomial in a small set of invariants, while at six points many perfectly valid bases make the same object look like a delicate rational function whose simplicity is only visible after cancellations that SR has no reason to guess early. Further improvements are therefore necessary to make progress at six points and beyond.

\section{Comparison with Neural Networks}
As we have seen, symbolic regression is a powerful tool for discovering analytic relations in scattering amplitudes, especially when the answer is \textit{simple}. In this sense, there is an obvious use-case for symbolic regression as a simplification tool for amplitudes: given an unwieldy expression for some amplitude, we can attempt to discover an equivalent analytic expression with a much lower complexity score. However, there are several other machine learning paradigms on the market \cite{Vaswani:2017lxt,Hochreiter:1997lstm,Gilmer:2017nmp,Huy:2026wcd}, and it is natural to ask how these might perform at the same task. Thankfully, there has been a recent attempt to do just this with transformer based neural networks, where a network is trained on pairs of equivalent expressions, one complex and one simple, with the goal of inferring simple expressions from complex ones~\cite{Cheung:2024svk}.

There is an immediate conceptual distinction between the two approaches. The transformer setup is a genuinely \emph{symbolic-to-symbolic} map: the input is an explicit expression (represented as a token sequence) and the output is another explicit expression, with the network implicitly learning which algebraic identities are useful and how to apply them. Symbolic regression, by contrast, is most naturally a \emph{numeric-to-symbolic} procedure: it is given numerical evaluations of a target function on phase space and attempts to infer a closed-form expression that reproduces those values. For amplitudes this difference is not cosmetic. The neural network is, in principle, sensitive to the detailed \emph{presentation} of the expression (how it is bracketed, which sub-expressions appear, etc.), whereas symbolic regression only “sees” the function defined by that expression through its numerical values. This makes the symbolic-regression problem closer in spirit to rediscovering an amplitude from something akin to a measurement (or a black-box evaluation), rather than performing algebraic rewriting in the usual sense.

To compare the two paradigms on the same physical target, we focus on the benchmark highlighted in~\cite{Cheung:2024svk}: the five-point amplitude for three scalars and two same-helicity gravitons. In that work, a sequential transformer-based network, trained on complex-simple pairs, reduces a 298-term spinor-helicity expression to a compact two-term result (their eq.~(4.7)),
\begin{equation}
M(\phi\phi\phi h^+h^+)
=
\frac{\langle 12\rangle\langle 13\rangle\langle 23\rangle}{\langle 24\rangle\langle 25\rangle\langle 45\rangle}
\left(
\frac{[14][35]}{\langle 14\rangle\langle 35\rangle}
-
\frac{[15][34]}{\langle 15\rangle\langle 34\rangle}
\right).
\label{eq:ppphh_compact}
\end{equation}
From the point of view of our pipeline, this is an ideal test case: the “raw” expression is extremely long, while the simplified answer consists of only two rational monomials.

\subsection{Regression setup and feature design}
Our symbolic-regression comparison is deliberately set up so that the algorithm does \emph{not} receive the simplified form in eq.~\eqref{eq:ppphh_compact}. Instead, we numerically evaluate the full 298-term expression on a set of randomly generated phase-space points, and ask symbolic regression to infer a compact analytic formula consistent with those evaluations. Concretely, we proceed as before and generate exact rational on-shell five-point kinematics in split signature $(2,2)$ using spinors, so that $\langle ij\rangle$ and $[ij]$ are independent real numbers. For each phase-space point we compute the full set of angle and square brackets and store these as features, together with the numerical value of the target 298-term amplitude. The data-driven approach has an important advantage over the neural network approach: any candidate expression returned by symbolic regression is immediately verified on held-out data, and since the objects involved are rational functions of spinor brackets, agreement on a sufficiently large number of generic phase-space points makes an incorrect expression incredibly unlikely. The neural network, by contrast, returns a \textit{conjectured} simplified expression whose correctness must still be validated by numerical testing to rule out hallucinations, which are a common issue in neural symbolic manipulation.

A naive symbolic-regression attempt would feed all brackets $\{\langle ij\rangle, [ij]\}$ directly into a generic rational function search space generated by $\{+,-,\times,/\}$. As we have seen, however, this is an unnecessarily hard search problem, since the search space is very large. Furthermore, if this is all we feed in, then we are expecting the algorithm to simultaneously discover both the correct little-group weights and the correct overall mass dimension, all the while navigating the vast number of algebraically equivalent expressions.

As before, then, we will use our knowledge of physics to inject mild priors to constrain the problem, enforcing little-group covariance and dimensional consistency. For the $\phi\phi\phi h^+h^+$ amplitude, the scalars carry zero little-group weight while each positive-helicity graviton carries weight $-2h_i = -4$. This means that any valid term in the amplitude must have net little-group exponents
\[
(0,0,0,-4,-4)
\]
across legs $(1,2,3,4,5)$, where each $\langle ij\rangle$ contributes $+1$ to legs $i$ and $j$, and each $[ij]$ contributes $-1$ to legs $i$ and $j$. Since the target is a rational function of brackets, we can enumerate candidate \emph{rational monomials} built from products of $\langle ij\rangle$ and $[ij]$ in numerator and denominator that satisfy these little-group constraints (subject to modest caps on the total number of bracket factors). Evaluated on data, these monomials form a design matrix $\Phi$. We then proceed as before: we apply CPQR to remove redundant features, leaving an independent (and typically better conditioned) feature set.

There is no reason a priori to assume that the simplest expression is linear in these features alone: as we have seen, spurious poles can be introduced in intermediate terms assuming they cancel in the final expression, for example. That being said, dimensional consistency and little-group covariance only allows for a very restricted set of terms, and at this order we can simply enumerate all possible permitted terms and add them to the feature set. 
Operationally, this allows us to restrict the symbolic-regression operator set to $\{+,-,\times\}$, effectively making it linear regression, since division is already built into the monomials, substantially shrinking the search space while still containing the true answer.

With this setup, symbolic regression recovers eq.~\eqref{eq:ppphh_compact}, or an alternative simple expression equal by schouten or momentum conservation, in $\cl{O}(10^2)$ seconds, matching the 298-term target to $\cl{O}(10^{-16})$ maximum relative error on a held-out test set. The rediscovered expression is algebraically equivalent to the two-term form reported in~\cite{Cheung:2024svk}, confirming that SR can simplify a lengthy spinor-helicity expression to its simplest known form without ever seeing that form as input, and without extensive training or a large corpus of examples.

\subsection{Similarities and differences between neural and symbolic regression}
This benchmark usefully highlights the different inherent biases of the neural and symbolic approaches.

On the neural side, the transformer is explicitly learning an \emph{algorithm for simplification}: it must internalize patterns of identity application (Schouten, momentum conservation, partial fractioning, etc.) and learn how to rewrite a given expression into a shorter but equivalent one. This is powerful when one wants an automated “algebra engine” that can act directly on an input expression, and it naturally supports an iterative approach, i.e. it can simplify long expressions by breaking it into smaller expressions and simplifying those first.

On the symbolic-regression side, we are not learning a rewrite algorithm at all. Instead, we exploit the fact that amplitudes are highly constrained functions, and we build a feature basis that respects those constraints from the outset. In cases like eq.~\eqref{eq:ppphh_compact}, where the final answer is extremely simple in a physically natural basis, symbolic regression is well-matched to the task: it does not need to discover \emph{how} to simplify the 298-term input, only \emph{what} function that input represents.

The limitations of both approaches are also complementary. Neural simplification can, in principle, improve with larger and richer training corpora, and can act directly on complicated symbolic inputs, however it can often hallucinate: it returns a conjectured expression whose correctness must still be validated (e.g.\ by numerical testing or symbolic checking). Symbolic regression produces an explicit analytic candidate which is immediately verifiable on held-out data, but its performance depends sharply on feature construction: without strong physics-guided bases, the search space quickly becomes a combinatorial nightmare, and the method deteriorates as the number of independent structures grows. This is the same obstruction we saw already in the KLT rediscovery problem at higher multiplicity, and even when the true answer is “simple”, it may only be simple in a basis that is hard to find without additional structure.

Taken together, these observations suggest a pragmatic division of labour. Neural models are attractive as general-purpose symbolic manipulators which may discover useful intermediate rearrangements, while symbolic regression is most effective as an equation-discovery and \emph{final compression} tool once a physically meaningful feature set has been identified. With this in mind, it would be interesting to explore a hybrid approach, where a neural simplification step is used to preprocess the input expression into a more compact form, not necessarily the \textit{most} compact form, which is then fed into symbolic regression for final compression. For even higher efficiency, the neural step could be used to suggest promising features for the symbolic regression stage, for example by identifying which combinations of brackets appear most frequently in the simplified expression, or by suggesting which orderings of gluon amplitudes are most likely to yield simple KLT relations. This could potentially combine the best of both worlds: the neural network can learn to apply algebraic identities to reduce the complexity of the expression, while symbolic regression can then identify a compact analytic form that matches the simplified numerical data.

\section{Conclusion and Outlook}
In this letter, we have presented a data-driven pipeline that rediscovers several textbook flagship analytic structures in scattering amplitudes: the Kleiss--Kuijf and BCJ relations, the Parke--Taylor formula, and the KLT double-copy, all using only numerical on-shell data, standard linear algebra (CPQR), and symbolic regression, with a few well-motivated theoretical priors. The method succeeds at four and five points, where the BCJ basis is small and the KLT kernel is a short polynomial, but encounters a genuine combinatorial barrier at six points and beyond. A comparison with the neural-network simplification approach of~\cite{Cheung:2024svk} shows that the two approaches are complementary: symbolic regression excels at \emph{discovering} compact expressions from numerical evaluations, while neural rewriting excels at \emph{simplifying} known symbolic inputs.

The most pernicious obstacle we have come across in this work is the combinatorial explosion at higher multiplicity, and perhaps the most natural way to circumvent it is by adopting a factorisation bootstrap: rather than learning the full amplitude in one shot, we could probe the target at phase-space points approaching a chosen channel $s_I\to 0$, extract the residue of the scaled quantity $s_I\,\cl M$ via SR, and assemble a candidate pole part $\cl M_{\mathrm{poles}}$. This would leave some remaining contact terms $\cl M_{\mathrm{contact}}=\cl M-\cl M_{\mathrm{poles}}$, but since these are free of physical poles they should be substantially simpler to learn by SR. Initial tests of this idea are promising, but approaching poles cleanly requires care to avoid collinear regions and spurious-pole cancellations, and we leave this idea to future work. Should this approach prove successful, a natural next step would be to explore the application of these methods to more complex amplitude sectors where less analytic structure is known, such as at NMHV at six points and above. It would also be interesting to explore a hybrid neural--symbolic pipeline, in which a transformer first reduces the complexity of the input expression, perhaps to several smaller expressions, which it then hands-off to symbolic regression performs the final compression, combining the complementary strengths discussed in the previous section, perhaps undergoing several passes. On the physics side, extension to loop level mean the appearance of transcendental functions (logarithms, polylogarithms), which would require enlarging the operator library, although the same CPQR-based feature selection and physics-informed dimensional constraints apply in principle. A simpler target, staying at tree-level, might be to try and learn the string theory KLT relations, a recent target for bootstrapping \cite{Chen:2023dcx}, which would also involve enlarging the operator library to include trigonometric functions, but would still be a rational function of the kinematic invariants times the bilinears, and ought to be simple enough to learn. While we have focussed on a particular scattering-amplitudes formulation of the double copy, symbolic regression has a long track record in learning equations of motion, largely because trajectory data providing direct access to the underlying dynamics are often easy to obtain. Several non-perturbative formulations of the double copy can, in fact, be expressed at the level of equations of motion \cite{Moynihan:2025vcs,Cheung:2022bvq,Moynihan:2021rwh,Cheung:2021zvb,Emond:2021lfy,Moynihan:2020ejh,Armstrong-Williams:2022gbm,Carrasco:2026ijt}, and it would be interesting to investigate whether symbolic regression could help explore these formulations, or even suggest new ones.

More generally, we believe that the combination of symbolic regression and neural network-based simplification has the potential to be a powerful tool for exploring the rich structure of scattering amplitudes and uncovering new insights into the underlying physics, and we look forward to investigating this in more detail in future work.\newpage
\section{Acknowledgments} 
It is a pleasure to thank Kymani Armstrong-Williams, David Berman, Miles Cranmer, Ed Hirst, David Kosower, Costis Papageorgakis, and Chris White for useful conversations and feedback. This work was supported by the Science and Technology Facilities Council (STFC) Consolidated Grant ST/X00063X/1 ``Amplitudes, Strings \& Duality". No new data were generated or analysed during this study. Some of the code used to generate the results in this paper is available at \href{https://github.com/nmoynihan/amplitudes-sr}{https://github.com/nmoynihan/amplitudes-sr}.
\bibliography{references}
\end{document}